\documentclass[12pt]{iopart}
\pdfoutput=1
\usepackage{iopams} 

\usepackage{graphicx}
\usepackage{dcolumn}
\usepackage{bm}

\usepackage{hyperref}
\begin{document}

\newcommand{\N}{\ensuremath{\mathbb{N}}}
\newcommand{\G}{\ensuremath{\mathcal{G}}}
\newcommand{\md}{\ensuremath{\delta}}
\newcommand{\Z}{\ensuremath{\mathbb{Z}}}
\newcommand{\R}{\ensuremath{\mathbb{R}}}
\newcommand{\E}{\ensuremath{\mathbb{E}}}
\renewcommand{\Pr}{\ensuremath{\mathbb{P}}}

\newcommand{\DC}{\ensuremath{\emph{DC}}}
\newcommand{\CC}{\ensuremath{\emph{CC}}}
\newcommand{\EC}{\ensuremath{\emph{EC}}}
\newcommand{\BC}{\ensuremath{\emph{BC}}}
\newcommand{\PC}{\ensuremath{\emph{PC}}}
\newcommand{\GED}{\ensuremath{\emph{GED}}}
\newcommand{\KC}{\ensuremath{\emph{KC}}}

\newcommand{\CGD}{\ensuremath{\emph{CGD}}}
\newcommand{\GoG}{\ensuremath{GoG}}
\newcommand{\GoGs}{\ensuremath{GoG} }

\newcommand{\todo}[1]{\textcolor{red}{TODO: #1}}
\newcommand{\gilles}[1]{\textcolor{blue!40!red}{Gilles: #1}}
\newcommand{\matthieu}[1]{\textcolor{green!40!blue}{Matthieu: #1}}
\newcommand{\stefan}[1]{\textcolor{blue!100!red}{Stefan: #1}}
\newcommand{\ya}[1]{\textbf{YA: #1}}

\newtheorem{theorem}{Theorem}
\newtheorem{corollary}[theorem]{Corollary}
\newtheorem{lemma}[theorem]{Lemma}
\newtheorem{claim}[theorem]{Claim}
\newtheorem{definition}[theorem]{Definition}
\newtheorem{example}[theorem]{Example}
\newtheorem{remark}{Remark}
\newtheorem{observation}{Observation}

\newtheorem{fact}[theorem]{Fact}

\title{The Many Faces of Graph Dynamics}

\author{Yvonne Anne Pignolet$^1$ \quad Matthieu Roy$^2$ \quad Stefan Schmid$^3$ \quad Gilles Tredan$^2$}
\address{$^1$ ABB Corporate Research, Switzerland \quad $^2$ LAAS-CNRS, France\\
 $^3$ Aalborg University, Denmark \& TU Berlin, Germany}%
\date{\today}

\begin{abstract}
The topological structure of complex networks has fascinated researchers for several 
decades,  resulting in the discovery of many 
universal properties
and reoccurring characteristics of
different kinds of networks. 
However, much less is known today about the
\emph{network dynamics}: indeed, complex networks in reality are not
static, but rather \emph{dynamically evolve over time}.

Our paper is motivated by the empirical observation that 
network evolution patterns seem far from random, but
\emph{exhibit structure}. Moreover, the specific patterns
appear to depend on the  network type, contradicting
the existence of a ``one fits it all'' model.
However, we still lack observables to quantify these intuitions, 
as well as metrics to compare graph evolutions.
Such observables and metrics  are needed 
for extrapolating or predicting evolutions, as well as for interpolating
graph evolutions.

To explore
the many faces of graph dynamics and to
quantify  temporal changes, this paper suggests to
build upon
the concept 
of centrality, a measure of node importance in a network.
In particular, we introduce the notion of \emph{centrality distance}, 
a natural similarity measure for two graphs which depends
on a given centrality, characterizing the graph type.
Intuitively, centrality distances reflect the extent to which
(non-anonymous) node roles are different or, in case of dynamic graphs, have
changed over time, 
between two graphs.

We evaluate the centrality distance
approach for five evolutionary models and seven real-world social and 
physical networks. 
Our results empirically show the usefulness of centrality
distances for characterizing graph dynamics 
compared to a null-model of random evolution,
and highlight the differences between the considered 
scenarios.
Interestingly, our approach allows us to 
compare the dynamics of very different networks,
in terms of scale and evolution speed.

~ 
\end{abstract}



\maketitle

%
%
%
%
%


\section{Introduction}
\label{sec:introduction}

How do real-world networks evolve with time? 
While empirical studies provide many 
intuitions and expectations, many questions remain open.
In particular, we lack
tools to characterize 
and quantitatively compare temporal graph dynamics.
In turn, such tools require good observables to
quantify 
the (temporal) relationships between networks.

In particular, the few network dynamics models 
that currently exist 
are often \emph{oblivious of the network type}. This is problematic,
as complex networks come in many different flavors, 
including social networks, biological networks, or physical networks. 
It seems highly unlikely that these very different graphs evolve in a similar
manner.

A natural prerequisite to measure evolutionary distances are good metrics to 
\emph{compare} graphs. 
The classic similarity measure for graphs is the \emph{Graph Edit Distance 
($\GED$)}~\cite{ged-survey}: the graph edit distance~$d_{\GED}(G_1,G_2)$ between
two graphs~$G_1$ and~$G_2$ is defined as the minimal number of \emph{graph edit 
operations} that are needed to transform~$G_1$ into~$G_2$. The specific set of 
allowed graph edit operations depends on the context, but typically includes 
node and link insertions and deletions.
While graph edit distance metrics play an important role in computer graphics
and are widely applied to pattern analysis and recognition, \GED{} is not 
well-suited for measuring similarities of networks in other 
contexts~\cite{faloutsos}: the set of graphs at a certain graph edit 
distance~$d$ from a given graph~$G$ exhibit very diverse characteristics and 
seem unrelated; being oblivious to semantics, the \GED{} does not capture any 
intrinsic structure typically found in  real-world networks.

A similarity measure that takes into account the inherent structure
of a graph may however have many important applications. A large body of work on
graph similarities focusing on a variety of use cases have been developed in the
past (see our discussion in Section~\ref{sec:relwork}). Depending on the 
context in which they are to be used, one or another is more suitable.
In particular,
we argue that graph similarities and graph distance measures are also an 
excellent tool
for the analysis, comparison and prediction of temporal network traces,
allowing us to answer questions such as:
\emph{Do these two networks have a common ancestor?} \emph{Are two evolution
patterns similar?} or \emph{What is a likely successor network for a given 
network?}
However, we argue that in terms of graph similarity measures, 
there is no panacea:
rather, graphs and their temporal patterns, come with many faces. 
Accordingly, we in this paper, propose to use a parametric, 
\emph{centrality-based approach} to measure graph similarities and distances, 
which in turn can be used to study the evoluation of networks.

More than one century ago, Camille Jordan introduced the first graph 
\emph{centrality} measure in his attempt to capture ``the center of a 
graph''. Since then the family of centrality measures has grown larger 
and is commonly employed in many graph-related studies. All major 
graph-processing libraries commonly export functionality for degree, closeness, 
betweenness, clustering, pagerank and eigenvector centralities.
In the context of static graphs, \emph{centralities} have proven to be 
a powerful tool to extract meaningful information on the structure of
the networks, and more precisely on the \emph{role} every participant (node) has
in the network. In social network analysis, centralities are widely used 
to measure the importance of nodes, e.g., to determine key players in social 
networks, or main actors in the propagation of diseases, etc. 

Today, there is no consensus on ``good" and ``bad" centralities: each centrality 
captures a particular angle of a node's topological role, some of which can be 
either crucial or insignificant, depending on the application. \emph{Am I important
because I have many friends, because I have important friends, or because 
without me, my friends could not communicate together?} The answer to this 
question is clearly context-dependent.

In this paper, we argue that the perceived quality of network similarities or 
distances measuring the difference between two networks depends on the focus
and application just as much. Instead of debating the advantages and 
disadvantages of a set of similarities and distances, we provide a framework
to apply them to characterize network evolution from different perspectives. 
In particular, we leverage centralities to provide a powerful tool to quantify 
network changes. The intuition is simple: to measure how a network evolves, 
we measure the change of the nodes' roles and importance in the network, by
leaving the responsibility to quantify node importance to centralities.

\paragraph{Our Contributions}
This paper is motivated by the observation that centralities
can be useful to study the 
dynamics of networks over time, taking into account the individual
roles of nodes (in contrast to, e.g., isomorphism-based measures, 
as they are used in the context of anonymous graphs), 
as well as the context and semantics (in contrast to, e.g.,
graph edit distances). 
In particular, we introduce the notion of
\emph{centrality distance}~$d_{C}(G_1,G_2)$ 
for two graphs~$G_1,G_2$, a graph similarity measure
based on a \emph{node centrality}~$C$.

We demonstrate the usefulness of our approach to identify and characterize
the different faces of graph dynamics. To this end, we 
study five generative graph models and
seven dynamic real world networks in more details. 
Our evaluation methodology comparing the quality of different similarity measures
 to a random baseline using data from actual graph evolutions, 
 may be of independent interest.

In particular, we demonstrate how centrality distances provide 
interesting insights into the 
structural evolution of these networks and show that actual evolutionary 
paths are far from being random. 
Moreover, we build upon the centrality distance concept
 to construct
  dynamic graph signatures. The intuition is simple: we measure the probability
  of an update to be considered as an outlier compared to a uniformly random
  evolution. This allows us to quantify the deviation of a given dynamic network
  from a purely random evolution (our null-model) of the same structure for a set of
	centrality distances. The signature consisting of the resulting deviation 
	values enables the comparison of different dynamisms on a fair basis,
	independently from  
	scale and sampling considerations.

\paragraph{Examples}
To motivate the need for tools to
analyse network evolution, we consider two simple 
examples.

\textbf{Example 1.} [Local/Global Scenario]
Con\-sider three graphs~$G_1$,~$G_2$,~$G_3$ over five nodes~$\{v_1,v_2,\ldots,v_5\}$:~$G_1$ is a line,
where~$v_i$ and~$v_{i+1}$ are connected;~$G_2$ is a cycle, i.e.,~$G_1$ with
an additional link~$\{v_1,v_5\}$; and~$G_3$ is~$G_1$ with an additional link~$\{v_2,v_4\}$.
In this example, we first observe that~$G_2$ and~$G_3$ have the same graph edit distance to~$G_1$:~$d_{\GED}(G_1,G_2)=d_{\GED}(G_1,G_3)=1$,
as they contain one additional edge. However, in a social network context, one would intuitively expect~$G_3$
to be closer to~$G_1$ than~$G_2$. For example, in a friendship network a short-range ``\emph{triadic closure}''~\cite{triadic} link may be more likely
to emerge than a long-range link: friends of friends may be more likely to become friends themselves in the future.
Moreover, more local changes are also
expected in mobile environments (e.g., under bounded human mobility and speed).
As we will see, the centrality distance concept of this paper can capture such 
differences.

\begin{figure}[hb]
\centering
\includegraphics[width=0.7\columnwidth]{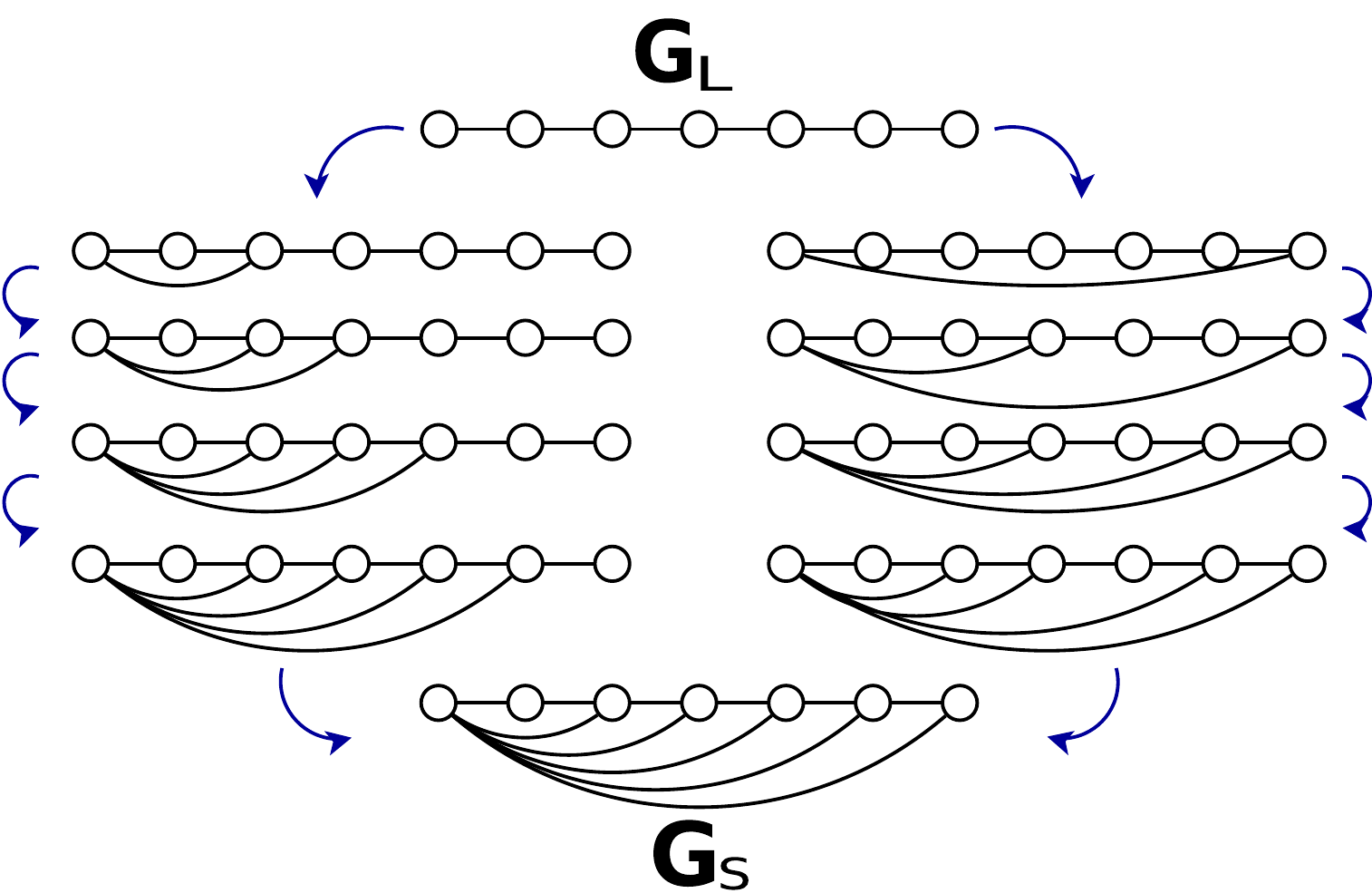}
\caption{
Two evolutionary paths from a line graph~$G_L$ to a shell
 graph~$G_S$.}
\label{fig:methodology}
\end{figure}

\textbf{Example 2.} [Evolution Scenario]
As a second synthetic example, consider two graphs~$G_L$ and~$G_S$,
where~$G_L$ is a line topology and~$G_S$ is a ``shell network''
(see also Figure~\ref{fig:methodology}).
How can we characterize evolutionary paths leading from the~$G_L$ topology to~$G_S$?
Note that the graph edit distance does not provide us with any information about 
the likelihood or the role changes of \emph{evolutionary paths} from~$G_L$ to 
$G_S$, i.e., on the order of  edge insertions: there are many possible orders 
in which the missing links can be added to~$G_L$, 
and these orders do not differ in any way when comparing them with the graph 
edit distance.
In reality, however, we often have some expectations on how a graph may have 
evolved between two given snapshots~$G_L$ and~$G_S$.
For example, applying the triadic closure principle to our example, we would 
expect that the missing links are introduced one-by-one, from left to right.

The situation may look different in technological, man-made networks. 
Adding links from left to right only slowly improves the ``routing 
efficiency'' of the network:
after the addition of~$t$ edges from left to right, the longest
shortest path is~$n-t$ hops, for~$t<n-1$. A more efficient evolution of the 
network 
is obtained by connecting $v_1$ to the furthest node, adding links to the middle of the network,
resulting in a faster distance reduction: after~$t$ edge insertions, the 
distance is roughly reduced by a factor~$t$.

Thus, different network evolution patterns can be observed in real networks.
Instead of defining application-dependent similarities with design choices 
focusing on which evolution patterns are more expected from a certain network, 
we provide a framework that allows the joint characterization of graph dynamics 
along different axes. 

\paragraph{Organization}
The remainder of this paper is organized as follows.
Section~\ref{sec:background} provides the reader with the necessary background.
Section~\ref{sec:graph-dist} introduces our centrality distance
framework and  Section~\ref{sec:methodology}
our methodology to study the different graph dynamics empirically.
Section~\ref{sec:experiments} reports on results from analyzing real and 
generated networks.  
After reviewing related work in Section~\ref{sec:relwork}, we conclude our
contribution in Section~\ref{sec:conclusion}.

\section{Preliminaries}\label{sec:background}

This paper considers \emph{labeled} graphs~$G=(V,E)$, where vertices
$v\in V$ have unique identifiers and are connected via undirected edges~$e\in E$.
In the following, we denote as~$\Gamma(v)$ the set of neighbors of node~$v$:
~$\Gamma(v)=\{ w\in V \textrm{~s.t.~}\{v,w\} \in E\}$.
A temporal network trace is a sequence~$T = \left[G_0, G_1, \ldots, G_l\right]$, 
 where~$G_i (V, E_i)$ represents the network at the~$i^{th}$ snapshot.

We focus on \emph{node centralities}, a centrality being a real-valued function assigning ``importance values''
to nodes. Obviously, the notion of importance is context-dependent, which has led 
to many different definitions of centralities. We refer to~\cite{borgatti2006graph} for a thorough and formal discussion on 
centralities.

\begin{definition}[Centrality]
 A \emph{centrality}~$C$ is a function~$C\colon (G,v) \to \mathbb{R}{+}$ that,
 given a graph~$G=(V,E)$ and a vertex~$v\in V(G)$, returns a non-negative value
~$C(G,v)$. The centrality function is defined over all vertices~$V(G)$ of a
 given graph~$G$. 
\end{definition}

By convention, we define the centrality of a node without edges to 
be~0. We write $C(G)$ to refer to the vector in $\mathbb{R}_+^{n}$ where the 
$i^{th}$ element is $C(G,v_i)$ for a given order of the identifiers.

Centralities are a common way to characterize networks and their vertices.
Frequently studied centralities include the \emph{degree
centrality} (\DC), the \emph{betweenness centrality} (\BC), the \emph{closeness centrality}
(\CC), and the \emph{pagerank centrality} (\PC) among many more. A node is~$\DC$-central if it has many edges: the degree centrality is simply the node degree;
a node is~$\BC$-central if it is on many shortest paths: the betweenness centrality is the number of shortest paths going
through the node; a node is~$\CC$-central if it is close to many other nodes: the closeness centrality
measures the inverse of the distances to all other nodes; and a node is~$\PC$-central if the probability that a random walk on $G$ visits this node is high.
We use the classical definitions for centralities, and the exact formulas are presented in~\ref{appendix:centrality} for the sake of completeness.

Finally, throughout this paper, we will define the graph edit distance~$\GED$
between two graphs~$G_1$ and~$G_2$ as the minimum number of operations to
transform~$G_1$ into~$G_2$ (or vice versa), where an \emph{operation} is one of
the following: link insertion and link removal. 

%

\section{Centrality Distance}\label{sec:graph-dist}

The canonical distance measure is the graph
edit distance,~$\GED$.
However,~$\GED$ often provides
limited insights into the graph dynamics in practice.
Figure~\ref{fig:methodology} shows 
an example with two evolutionary paths: an incremental~(\emph{left}) and
a binary~(\emph{right}) path that go from~$G_L$ to~$G_S$. With respect to~$\GED$, 
there are many equivalent shortest paths for moving from 
$G_L$ to~$G_S$. However, intuitively, not all traces are equally likely 
for dynamic networks, as the structural roles that nodes in networks have are 
often preserved and do not change arbitrarily. 
Clearly, studying graph evolution with GED thus cannot help us to understand how
structural properties of graphs evolve.

\begin{observation}
The graph edit distance~$\GED$ does not provide much insights into
graph evolution.
\end{observation}

We in this paper aim to enrich the graph similarity measure
with semantics. At the heart of our approach lies the concept of
\emph{centrality distance}: a simple and flexible
tool to study the similarity of graphs. 
Essentially, the centrality distance measures
the similarity between two centrality vectors.
It can be used to measure the distance between
two arbitrary graphs, not only between graphs
with graph edit distance 1.

\begin{definition}[Centrality Distance]
 Given a centrality~$C$, we define the centrality distance~$d_C(G_1,G_2)$
 between any two graphs as the sum of the node-wise difference of the centrality values:
~$$
d_C(G_1,G_2)=||C(G_1)-C_2(G_2)||_1=\sum_{v\in V}|C(G_1,v)-C(G_2,v)|
~$$
\end{definition}

Thus, the centrality distance intuitively measures the magnitude 
by which the roles of different nodes change. 
While we focus on the 1-norm in this paper,
the concept of centrality distance can be useful also for other norms.

Both the importance of node roles as well as the importance of
  node role changes is application-dependent. 
  Due to the large variety of processes dynamic
  graphs can capture, there is no one-size-fits-it-all measure
  of importance. 
  To illustrate this point, let us consider the  
  ``intuitive'' similarity properties
  proposed by Faloutsos et al.~\cite{deltacon}.
	For instance, the proposed 
	edge importance property should
	penalize changes that create disconnected components more than changes 
	that maintain the connectivity properties of the graphs.
 Now imagine a cycle graph of 100 nodes $c_1,..,c_{100}$,
  and a single additional node $v$ connected to $c_1$. According
  to the proposed edge importance property the most important link is $(c_1,v)$. Indeed, it
  is the only link whose removal would create a disconnected component
  (containing $v$ alone). Yet the removal of any other link would double the
  diameter of the structure. Or in an information dissemination network all 
	nodes would have to update half of their
  routing tables. So which link is more important? The answer clearly
  depends on the context.
Similar examples can be found for other properties 
proposed in~\cite{deltacon}, e.g., 
  regarding submodularity and focus-awareness. Not only are these
  properties hard to 
	formalize, their utility varies from application to application.

We conclude by noting
that given two centralities $C_1$ and $C_2$ and two arbitrary graphs $G_1$ and $G_2$
with $n$ nodes, the respective distances are typically different, i.e., 
$d_{C_1}(G_1, G_2) \neq d_{C_2}(G_1, G_2)$. Hence, using a set of different 
centrality distances, we can explore the variation of the graph dynamics 
in more than one ``dimension''.

\section{Methodology}\label{sec:methodology}

In order to characterize the different faces of graph dynamics
and to study the benefits of centrality-based measures,
we propose a simple methodology.
Intuitively, given a centrality capturing well the roles of different nodes
in a real-world setting, we expect the centrality distance
between two consecutive graph snapshots 
$G_t$ and~$G_{t+1}$ to be smaller than the typical distance 
from~$G_t$ to other graphs that have the same GED.

To verify this intuition, we define a 
{\em null model} for evolution. A null model generates networks using patterns 
and randomization, i.e., 
certain elements are held constant and others are allowed to 
vary stochastically. Ideally, 
the randomization is designed to mimic the outcome of a random process that would be 
expected in the absence of a particular 
mechanism~\cite{gotelli1996null}. 
Applied to our case, this means that starting from a given snapshot $G_t$
that represents the fixed part of the null model, 
if the evolution follows a null model, then any graph randomly
generated from $G_t$ at the given $\GED$ is evenly likely to appear. 

Concretely, for all consecutive
graph pairs~$G_t$ and~$G_{t+1}$ of a network trace, 
we determine the graph edit
distance (or ``radius'')~$R=d_{\GED}(G_t,G_{t+1})$.  Then, we generate a set
$S_{t+1}$ of~$k=100$ sample graphs~$(H_i)_{i=1..k}$ at the same \GED~$R$ from
$G_t$ \emph{uniformly at random}. That is, to create $H_i$,  we first start from a copy of 
$G_t$ and select $R$ node 
pairs, $(u_l, w_l)\in V^2, 1\leq l \leq R$, uniformly at random. For each of these pairs $(u_l, w_l)$ we add the edge 
$(u_l, w_l)$ to $H_i$ if it does not exist in $G_t$ or we remove it if it was in 
$G_t$ originally.  Such randomly built sample graphs at the same graph edit 
distance allow us to assess the impact of a uniformly random evolution of the 
same magnitude from the
same starting graph~$G_t$: $\forall H_i \in S_{t+1},
d_{\GED}(G_t,H_i)=d_{\GED}(G_t,G_{t+1}).$ In other words, $G_t$ is the 
pattern and the evolution to $H_i$ at
graph edit distance $R$ is the randomized
part of the null model\footnote{This is the least constrained randomization of
 network evolution w.r.t. the graph edit distance. More refined null
models may preserve other 
structural graph properties in the sample graphs, e.g., their densities.~\ref{appendix:altNullModel}
describes results obtained for a null model that guarantees the  average degree of $G_{t+1}$ in the sample graphs.}.

As a next step, given a centrality~$C$, we compare $G_{t+1}$ with the set $S_{t+1}$
that samples the evolution following the null model. We consider that $G_{t+1}$
does not follow the null model if it is an {\em outlier} in the set $S_{t+1}$
for the centrality $C$. Practically,
$G_{t+1}$ is considered an outlier if the absolute value of its distance from 
$G_t$ minus the mean distance of $S_{t+1}$ to $G_t$ is at least
twice  the standard deviation, 
i.e., if 
$$ | d_C(G_t,G_{t+1})-~\mu(\{d_C(G_t,x),x\in S_{t+1}\})| ~> 
					2\sigma(\{d_C(G_t,x),x\in S_{t+1}\nonumber \}).$$

Given a temporal trace $T$, we define
$p_{C,T}$ as the fraction of outliers in the trace for centrality $C$.
An ensemble of such values $p_{C_i,T}$ for a set of centralities $\mathcal{C}=\{C_1, \ldots, C_k\}$ is called a \emph{dynamic signature} of $T$.                                

\section{Experimental Case Studies}\label{sec:experiments}

Based on our centrality framework and methodology, 
we can now shed some light on the different faces of 
graph dynamics, using real world 
data sets.

\begin{itemize}
\item \textbf{Caida (AS):} This data captures the Autonomous Systems relationships as
 captured by the Caida project. Each of the~$400$ snapshots represents the
 daily interactions of the~$1000$ first AS identifiers from August 1997 until December
 1998~\cite{konect}.
\item \textbf{ICDCS (ICDCS):} We extracted the most prolific  authors in the ICDCS conference (IEEE International Conference on Distributed Computing Systems) and the 
co-author graph they form from the DBLP publication database (\url{http://dblp.uni-trier.de}). This trace contains 33 snapshots of 691 nodes
and 1076 collaboration edges. The timestamp assigned to an edge
corresponds to the first ICDCS paper the authors wrote together.
Clearly, the co-authorship graph is characterized by a strictly monotonic densification
over time.
\item  \textbf{UCI Social network (UCI):} The third case study is based on a publicly available
 dataset~\cite{opsahl2009clustering}, capturing all the messages exchanges
 realized on an online Facebook-like social network between 1882
 students at University of California, Irvine over 7 months. We discretized
 the data into a dynamic graph of 187 time steps representing the daily message
 exchanges among users.

\item  \textbf{Hypertext (HT):} Face-to-face interactions of the ACM Hypertext 2009 conference
 attendees. 113 participants were equipped with RFID tags. Each snapshot represents
 one hour of interactions~\cite{konect}.

\item  \textbf{Infectious (IN):} Face-to-face interactions of the ``Infectious: Stay away''
 exhibition held in 2009. 410 Participants were equipped with RFID tags. Each snapshot represents
 5 minutes of the busiest exhibition day~\cite{konect}.

\item  \textbf{Manufacture (MA):} Daily internal email exchange network of a medium-size
 manufacturing company (167 nodes) over 9 months of 2010~\cite{konect}.
 
\item  \textbf{Souk (SK):} This dataset captures the social interactions of 45
individuals during a cocktail, see~\cite{souk} for more details.
The dataset consists of~$300$ snapshots, describing the dynamic interaction graph between the participants, one
time step every~$3$ seconds~\cite{souk}.
\end{itemize}

\begin{figure}[!hbt]
	\centering
		\includegraphics[width=0.7\textwidth]{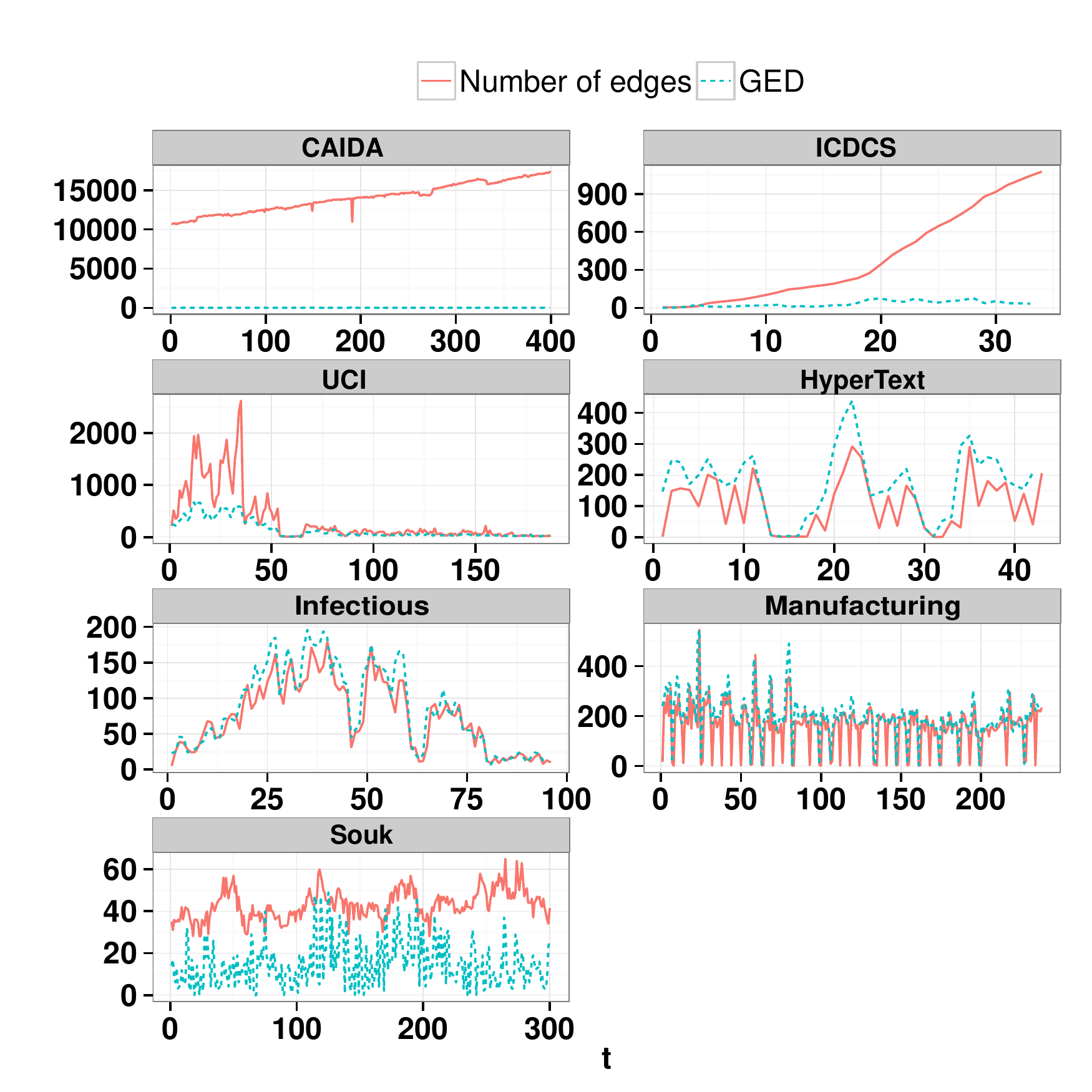} \vspace{-.5cm}
	\caption{Number of edges and graph edit distance $\GED$ in the network traces.}
	\label{fig:datasets}
\end{figure}

Figure~\ref{fig:datasets} provides a temporal overview on the evolution of the
number of edges in the network and the \GED{} between consecutive snapshots. 
Some of the seven datasets exhibit very different dynamics:
one can observe the time-of-day effect of attendees interactions on
\emph{Hypertext}, and the day-of-week effect on \emph{Manufacture}.  \emph{UCI},
\emph{Hypertext}, \emph{Infectious} and \emph{Manufacture} all exhibit a high
level of dynamics with respect to their number of links. This is expected for
 \emph{Infectious}, as visitors come and leave regularly and
rarely stay for long, but rather surprising for \emph{Manufacture}.

The density 
of \emph{Caida} slowly increases, and with a steady \GED. Similarly, the
number of co-author edges of \emph{ICDCS} steadily increases over the years, 
while the number of new edges per year is relatively stable. The number of days 
of the conference \emph{Hypertext} and the fact that conference participants 
sleep during the night and do not engage in social activity is evident in the 
second trace. The dynamic pattern of the online social network \emph{UCI} 
has two regimes: it has a high dynamics for the first 50
timestamps, and is then relatively stable, whereas \emph{Souk} exhibits a more
regular dynamics.
Generally, note that \GED{} can be at most twice as high as the maximal edge count
of two consecutive snapshots.

\subsection{Centrality Distances over Time}

\begin{figure}[htb]
	\centering
%
	\includegraphics[width=.33\textwidth]{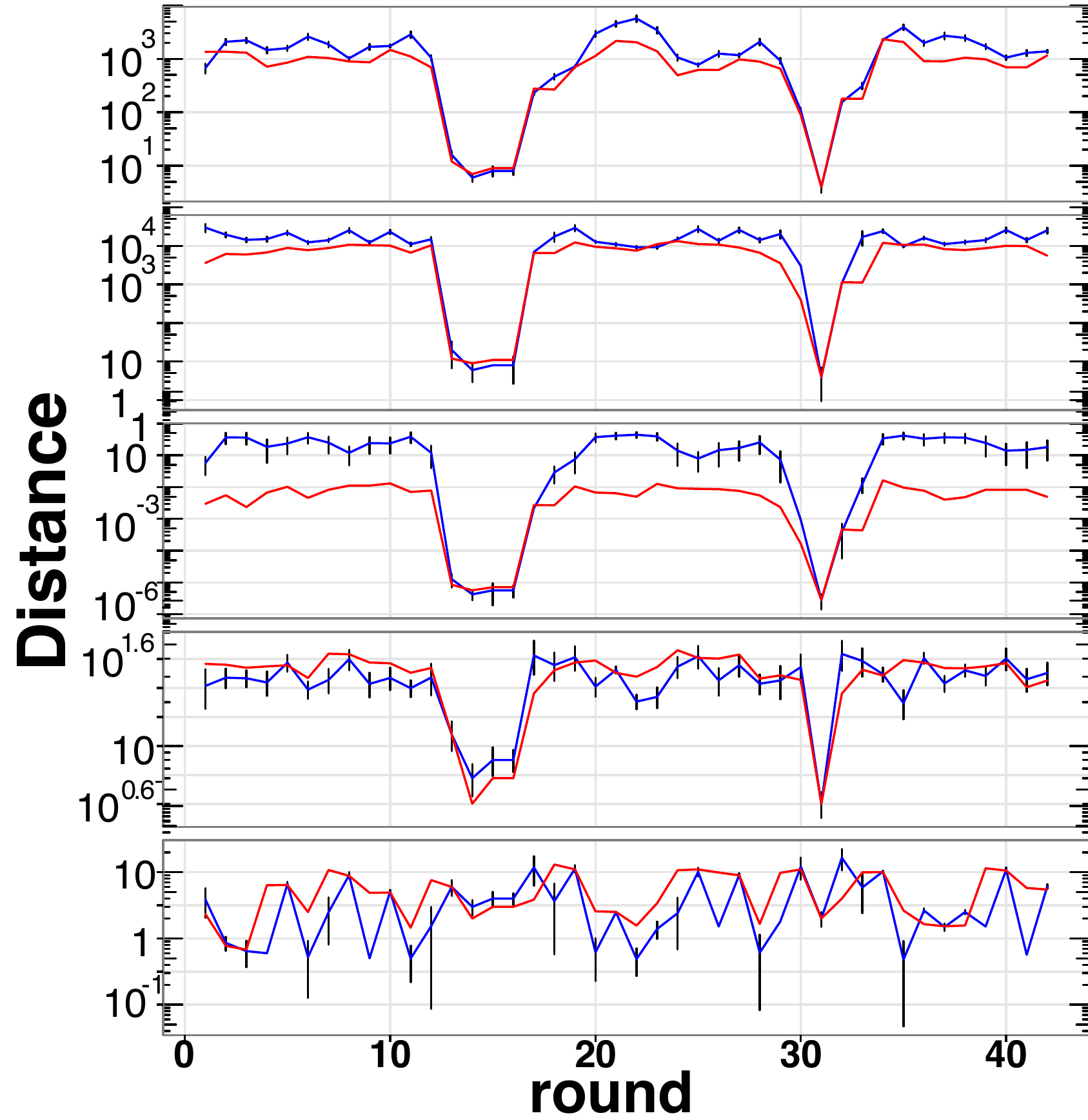}
	\includegraphics[width=.31\textwidth]{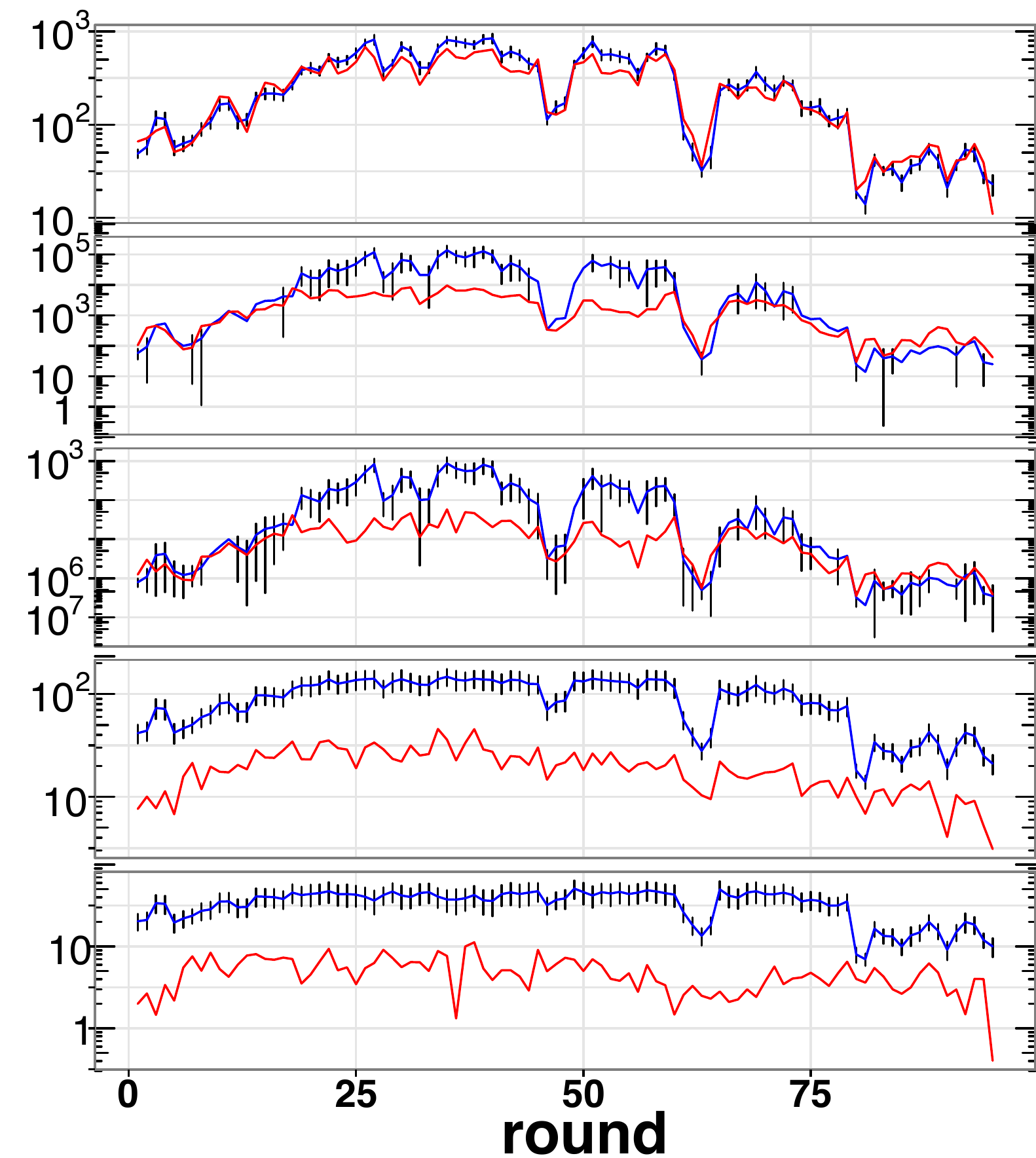}
		\includegraphics[width=.34\textwidth]{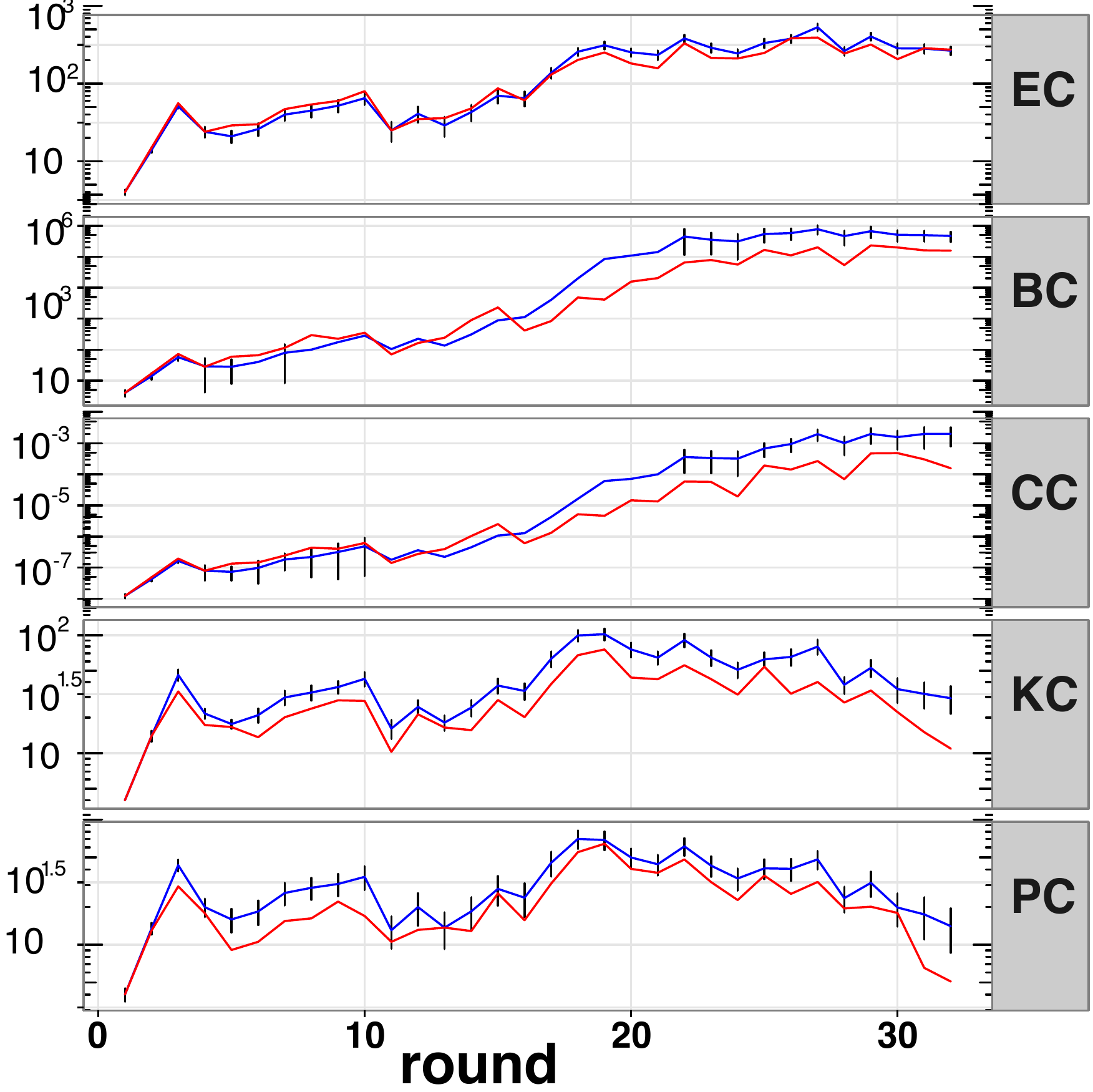}
\\
	\footnotesize\hfill Hypertext \hfill\hfill Infectious \hfill \hfill ICDCS \hfill
	\vspace{-.2cm}
	\caption{Centrality distance between~$G_t$ and~$G_{t+1}$ in dashed red lines
	and between~$G_t$ and 100 graphs with the same GED as~$G_t$ and~$G_{t+1}$ in
	solid blue lines representing the median, $2\sigma$ bars in grey. {\sf EC}: Ego centrality, {\sf BC}: Betweenness centrality, {\sf CC}: Closeness centrality, {\sf KC}: Cluster centrality, {\sf PC}: Pagerank centrality.}
	\label{fig:Chronogram}	
\end{figure}

Figure~\ref{fig:Chronogram}  presents examples of the results of our comparison 
of random graphs with the same graph edit distance GED as real-world network 
traces. The red dashed lines represent the centrality distances of 
$G_t$ and~$G_{t+1}$. The distribution of~$d_C$ values from
$G_t$ to the~$100$ randomly sampled graphs of~$S_{t+1}$ is represented as
follows: the blue line is the median, while the gray lines represent
the~$2\sigma$ outlier detection window.

For most graphs under investigation and for most centralities it holds 
that the induced centrality distance between~$G_t$ and~$G_{t+1}$ is often lower 
than between~$G_t$ and an arbitrary other graph with the same distance. There 
are however a few noteworthy details.

\emph{Hypertext} and \emph{Infectious} exhibit very similar dynamics compared from a
  GED perspective as shown in Figure~\ref{fig:datasets}. Yet from the other
  centralities' perspective, their dynamism is very different. Consider for
  instance \emph{Infectious} for
  \PC, where the measured distance is consistently an order of magnitude less
  than the sampled one. This can be understood from the link creation mechanics:
in \emph{Infectious}, visitors at different time periods never meet. By connecting these in principle very remote 
  visitors, the null model
  dynamics creates highly important links. This does not happen in \emph{Hypertext}
  where the same group of researchers meet repeatedly.
	In the monotonically growing co-authorship network of ICDCS, we can 
	observe that closeness and (ego) betweenness distances grow over time, which is not the case for the other networks in Figure~\ref{fig:Chronogram}.

       When looking at other centrality distances, we observe that even though
        the local structure changes, a different set of properties remains mostly
        unaltered across different networks. Moreover, for some 
        \emph{(graph, distance)}
        pairs, like~$\KC$ on \emph{ICDCS},~$\CC$ on \emph{Hypertext}, or~$\PC$ on
        \emph{Infectious}, the measured distance is orders of magnitude lower than the
        median of the sampled ones. This underlines a clear difference between 
				random
        evolution and the observed one from this centrality perspective: the
        link update dynamics is biased.

\subsection{Dynamics Signature}		
\begin{figure}[t]
	\centering
		\includegraphics[width=.6\textwidth]{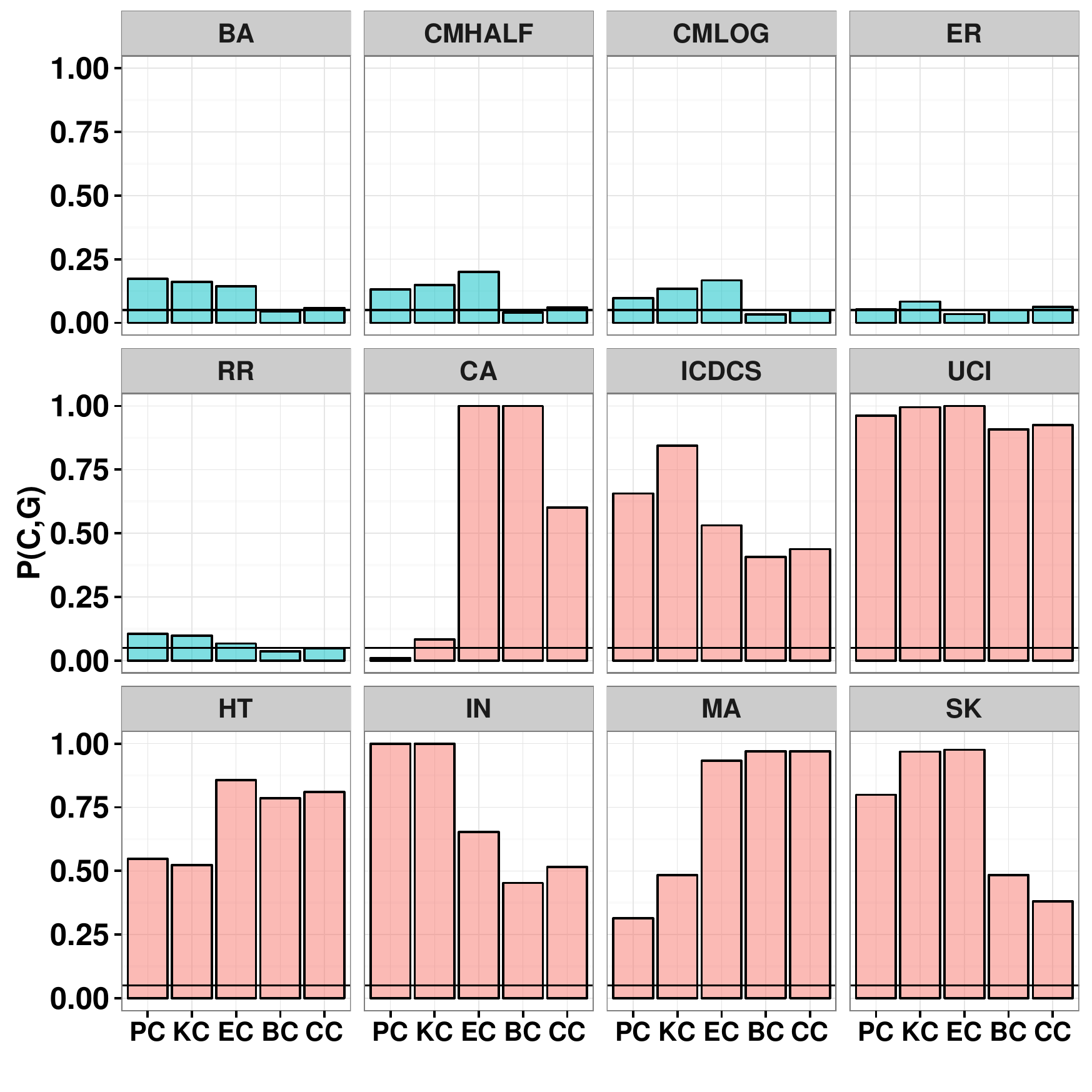}
	\caption{Histogram representation of the different facets
	of graph dynamics, the {\em dynamics signatures}. 
	For each data set, a histogram chart for the different 
	centralities is depicted, showing $p_{C,T}$, the probability that $G_{t+1}$ is
	an outlier w.r.t.~the
	null model for the corresponding centrality. Synthetic scenarios are depicted
	in \emph{blue}, real scenarios in \emph{red}. The black line at $5\%$ represents
	the null model, i.e., the fraction of graphs that are at distance at least $2\sigma$ from the mean in
	a normal distribution.  
	{\sf Synthetic datasets:} 
	{\sf BA}: Barabasi-Albert, {\sf CMHALF}: Preferential attachment---equiprobable nodes and edge events {\sf CMLOG}: Preferential attachement---node events decay in log, {\sf ER}: Erd\"os-R\'enyi, {\sf RR}: Random Regular.
	{\sf Real-life datasets:} {\sf CA}: Caida, {\sf ICDCS}: ICDCS co-authors,
	{\sf UCI}: Online social network of UCI, {\sf HT}: Hypertext conference, {\sf IN}: Infectious {\sf MA}: Manufacture mails, {\sf SK}: Souk cocktail.}
		\label{fig:spider}
	\vspace{-.2cm}
	\end{figure}

Figure~\ref{fig:spider} summarizes the~$p_{C,G}$ signatures for~$C\in \{
\CC,\EC,\BC,\PC,\KC\}$ applied to 7 real and 5 synthetic graphs in the form of a histogram chart---for synthetic
  graphs, each point is the average of 50 independent realizations of the model,
  and~$\vert S_t\vert=100$. That is, each
chart represents the probability of having graph evolutions being outliers with
respect to the null model for the corresponding centralities.  Interestingly,
this ``distinction ratio'' is not uniform among datasets. On \emph{Caida},
\emph{Infectious} and \emph{UCI}, the ratio is high for local centralities such
as \emph{PageRank} and \emph{Clustering}, and low for global centralities such
as \emph{Closeness} or \emph{Betweenness}. On the contrary, \emph{Hypertext} and
\emph{Manufacture} exhibit large ratios for global centralities and small ratios
for local centralities. Both local and global centralities perform well on Souk.
The difference of these behaviors show that these graphs adhere to different
types of dynamics.

To complement our observations on real networks with graph snapshots produced
according to a model, we investigated graph traces generated by some of the most
well-known models: \emph{Erd\H{o}s-R{\'e}nyi} ER~\cite{erd6s1960evolution},
\emph{random regular}
RR~\cite{steger1999generating}, \emph{Barabasi-Albert} BA~\cite{bar-alb}  and
\emph{preferential attachment}~\cite{chung2006complex} graphs with an equal
number of node and edge events (CMHALF) and with the number of node events
depending logarithmically on the time (CMLOG).  Perhaps the most striking
observation is that all tested dynamic network models have low~$p_{C,T}$ values
for all~$C$. This is partly due to the fact that the graph edit distance between
two subsequent snapshots is one and thus the centrality vectors do not vary as
much as between the snapshots and the sampled graphs of the same graph edit
distance for the real networks. Moreover, 
these randomized
  synthetic models are closer to the null model, and lack some of the
  characteristics (like link locality) of real world networks.
Furthermore, we observe that each random network model exhibits distinct 
dynamics signatures, with ER being closest to the null model.

\section{Related Work}
\label{sec:relwork}

To the best of our knowledge, our paper is the first to combine
the concepts of centralities and graph distances.
In the following, we review related work in the two
fields in turn, and subsequently discuss additional literature on dynamic 
graphs.

\textbf{Graph characterizations and centralities.}
Graph structures are often characterized by the frequency of small patterns
called \emph{motifs}~\cite{motif,motifs,Schreiber05frequencyconcepts,motifs-bioinf}, also known as \emph{graphlets}~\cite{graphlet},
or \emph{structural signatures}~\cite{signatures}.
Another important graph characterization, which is studied in this paper,
are \emph{centralities}~\cite{brandes}.
Dozens of different centrality indices have been defined over the last years,
and their study is still ongoing, with no 
unified theory yet. We believe that our centrality distance
framework can provide new inputs for this discussion.

\textbf{Graph similarities and distances.}
Graph edit distances have been used extensively in the context of inexact graph 
matchings in the field of pattern analysis. 
We refer the reader to the good
survey by Gao et al.~\cite{ged-survey}.
Soundarajan et al~\cite{soundarajan2014guide} compare twenty network 
similarities for anonymous networks. They distinguish between comparison levels 
(node, community, network level) and identify vector-based, classifier-based,
and matching-based methods. Surprisingly they are able to show that the results 
of many methods are highly correlated. 
NetSimile~\cite{netsimile} allows to assess the similarity between 
$k$ networks, possibly with different sizes and no overlaps in nodes or links.
NetSimile uses different social theories to compute similarity scores that are 
size-invariant, enabling mining tasks such as clustering, visualization, 
discontinuity detection, network transfer learning, and re-identification across
networks.
The Deltacon method~\cite{deltacon} is based on the normed difference of node-to-node 
affinity according to a Belief Propagation method. More precisely, the 
similarity between two graphs is the Root Euclidean Distance of their two 
affinity matrices or an approximation thereof.
The authors provide three axioms that similarities should satisfy and 
demonstrate using examples and simulations that their similarity features the desired 
properties of graph similarity functions.
Our work can be understood as an attempt to generalize
the interesting approach by Faloutsos et al.~in~\cite{deltacon},
which 
derives a distance from a normed matrix difference, where each element depends on
the relationships among the nodes. In particular, are argue that
there is no one-size-fits-it-all measure, and propose an approach
parametrized by centralities. Interestingly, we also prove that 
distances derived in our framework satisfy the 
axioms postulated in~\cite{deltacon}.

\textbf{Dynamic graphs.} 
Among the most well-known evolutionary patterns are the shrinking diameter
and densification~\cite{leskovec2007graph}. A lot of recent work studies 
link prediction 
algorithms~\cite{allali2013internal,linkprediction,yang:friendship}. 
Others focus on methods for finding frequent, coherent or 
dense temporal 
structures~\cite{jin2005discovering,shah2015timecrunch,sun2007graphscope},
or the evolution of communities and user behavior~\cite{ferlez2008monitoring,zhao2012multi}.

Another line of research attempts to extend the concept of centralities to dynamic graphs~\cite{costa2015time,kim2012temporal,lerman2010centrality,tabirca2011snapshot}. 
Some researchers study how the importance of nodes changes over time in dynamic networks~\cite{tabirca2011snapshot}. Others define temporal centralities which to rank nodes in dynamic networks and study their distribution over time~\cite{kim2012temporal,lerman2010centrality}. Time centralities which describe the relative importance of time instants in dynamic networks are proposed in~\cite{costa2015time}. 
In contrast to this existing body of work, our goal is to facilitate the direct comparison 
of entire 
networks and their dynamics, not only parts thereof.

A closely related work but using a different approach is by 
 Kunegis~\cite{kunegis}. Kunegis studies the evolution of networks 
 from a spectral graph theory perspective.
He argues that the graph spectrum describes a network on the global level, whereas eigenvectors describe a network at the local level,
and uses these results to devise link prediction algorithms.

\textbf{Bibliographic note.}
An early version of this work appeared at the ACM FOMC 2013 workshop~\cite{fomc14}.

\section{Conclusion}\label{sec:conclusion}

This paper was motivated by the observation that in terms of graph similarity measures,
there is no ``one size fits it all''. In particular, we have proposed a centrality-based
distance measure, and introduced a simple methodology to study the different faces
of graph dynamics.
Indeed, our experiments confirm that the evolution patterns of 
dynamic networks are not universal, and different networks need different centrality
distances to describe their behavior.
We observe that the edges in networks represent structural characteristics that are inherently 
connected to the roles of the nodes in these networks. These structures are 
maintained under changes, which explains the inertia of centrality distance 
which capture these properties. 
This behavior can be used to distinguish between natural and random network 
evolution. After analyzing a temporal network trace with a set of distance 
centralities, one can guess with confidence for future snapshots if they belong 
to the trace.

We believe that our work opens a rich field for future research. In this paper,
we focused on five well-known centralities and their induced distances, and
showed that they feature interesting properties when applied to the use case of
dynamic social networks. However, we regard our approach 
as a \emph{similarity framework},
which can be configured with various
 additional centralities 
and metrics, which may not even be restricted by
distance metrics, but can be based on the angles
between centrality vectors or use existing
correlation metrics (e.g., 
Pearson correlation, Tanimoto coefficient, log likelihood).
Finally, exploiting the properties of centrality distances,
especially their ability to distinguish and quantify between similar 
evolutionary traces, also opens the door to new applications, such as graph 
interpolation (what is a likely graph sequence between two given snapshots of a 
trace) and extrapolation, i.e., for link prediction algorithms based on 
centralities. 

\section{Acknowledgements}
The authors would like to thank Cl\'ement Sire for insightful remarks
on a previous version of this document. 

\begin{appendix}

\section{Centrality Definitions}
\label{appendix:centrality}

\noindent {\textbf{Degree Centrality $\DC$:}} Recall that
~$\Gamma(v)$ is the set of neighbors of a node~$v$. 
 The \emph{degree centrality} is defined as:~$$\DC(G,v)=\vert \Gamma(v) \vert.$$

\noindent
 \textbf{Betweenness Centrality $\BC$:}
 Given a pair~$(v,w) \in V(G)^2$, let~$\sigma(v,w)$ be the number of
 shortest paths between~$v$ and~$w$, and~$\sigma_x(v,w)$ be the number of shortest paths
 between~$v$ and~$w$ that pass through~$x \in V$. The {\it betweenness centrality}
 is:~$$\BC(G,v)=\sum_{x,w \in V}
 \sigma_v(x,w)/\sigma(x,w).$$ 
For consistency reasons, we consider that a node is on its own shortest path, i.e., 
~$\sigma_v(v,w)/\sigma(v,w)=1$, and, by
convention,~$\sigma_v(v,v)/\sigma(v,v)=0$. If $G$ is not connected, each
connected component is treated independently ($\sigma(x,w)=0\Rightarrow \forall
v, \sigma_v(x,w)/\sigma(x,w)=0 $). 

\noindent
 \textbf{Ego Centrality $\EC$:} Let~$G_v$ be the subgraph
 of~$G$ induced by~$(\Gamma(v) \cup \{v\})$. The \emph{ego centrality} is:
 $${EC}(G,v)={BC}(G_v,v).$$
 
\noindent
 \textbf{Closeness Centrality $\CC$:} Let~$\delta_G(a,b)$ be the length of a shortest path between vertices~$a$ and~$b$ in~$G$.
 The \emph{closeness centrality} is defined as:~$$\CC(G,v)=\sum_{w\in V\setminus v} 2^{-\delta_G(v,w)}.$$
	
\noindent
 \textbf{Pagerank Centrality $\PC$:} Let~$0<\alpha<1$ be a damping factor (e.g., the probability that a random person clicks on a link~\cite{page1999pagerank}). 
 The \emph{pagerank centrality} of $G$ is defined as: 
~$$\PC(G,v)=\frac{1-\alpha}{n}+\alpha\sum_{w\in V\setminus v} \frac{\PC(G,w)}{|\Gamma (w)|}.$$

\noindent
 \textbf{Cluster Centrality $\KC$}: 
 The \emph{cluster centrality} of a
 node~$v$ is the cluster coefficient
 of~$v$, i.e., the number of triangles in which~$v$ is involved
 divided by all possible triangles in~$v$'s neighborhood. By convention,~$\KC(G,v)=0$ for~$|\Gamma(v)|=0$, and~$\KC(G,v)=1$ for~$|\Gamma(v)|=1$. For higher degrees:$$\KC(G,v)=\frac{2\vert\{ \{j,k\}\textrm{~s.t.~} (j,k)\in \Gamma(v)^2, (j,k) \in E\}\vert}{|\Gamma(v)|(|\Gamma(v)|-1)}.$$

\section{Alternate Null Model Preserving Average Degree}
\label{appendix:altNullModel}

We present here some additional results related to an alternative choice of the 
null model. As described in the article,
we base our methodology on a uniformly random evolutionary null model that is 
based on the graph edit distance and hence may
not preserve some of intrinsic characteristics of networks under study, such as 
their density.

\begin{figure}[!h]
\centering
\includegraphics[width=.7\linewidth]{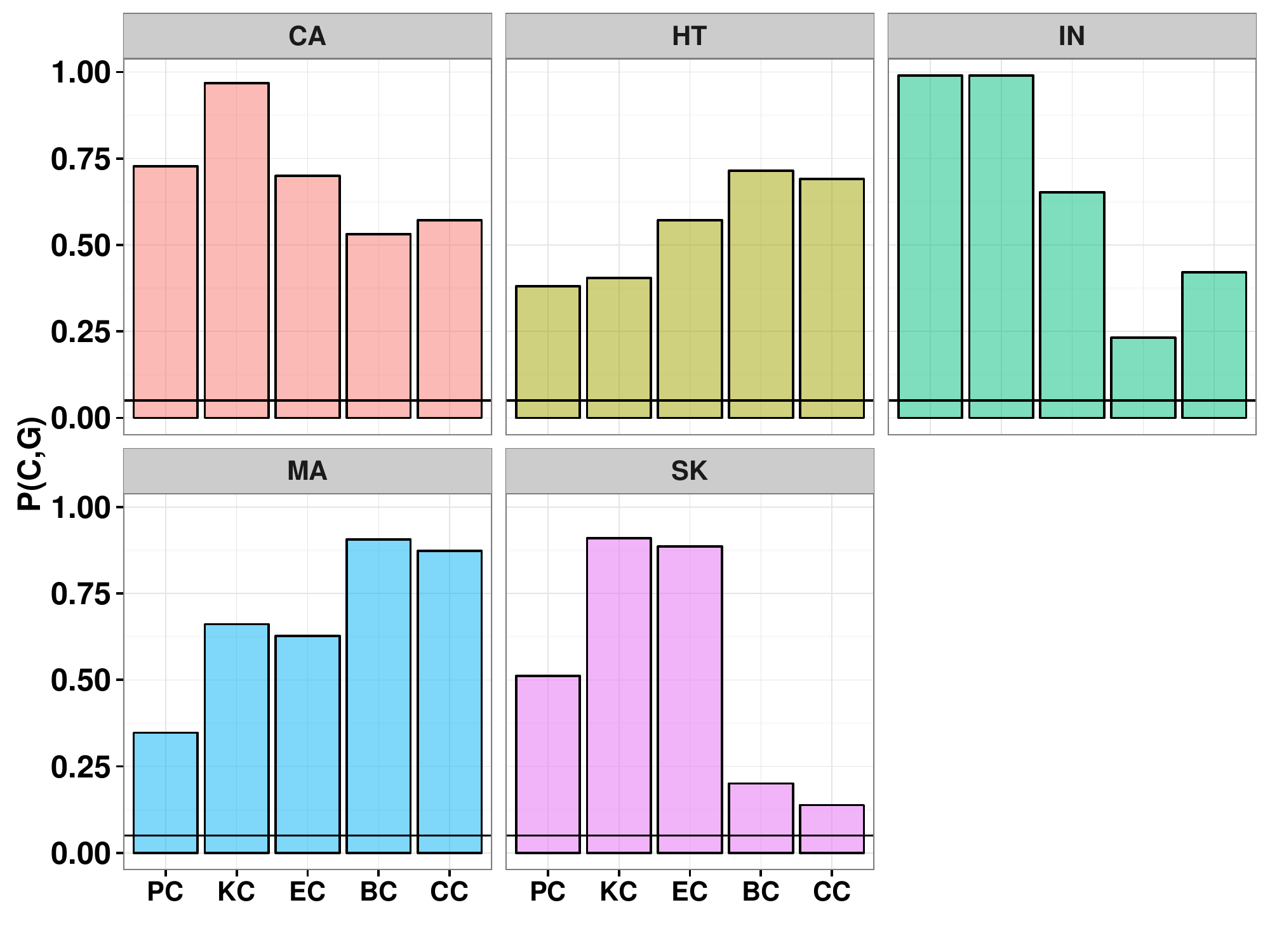}
\caption{Results of the proposed methodology using a null model
that preserves average degree}
\label{figure:altNull}
\end{figure}

\begin{figure}[!h]
\centering
\includegraphics[width=.6\linewidth]{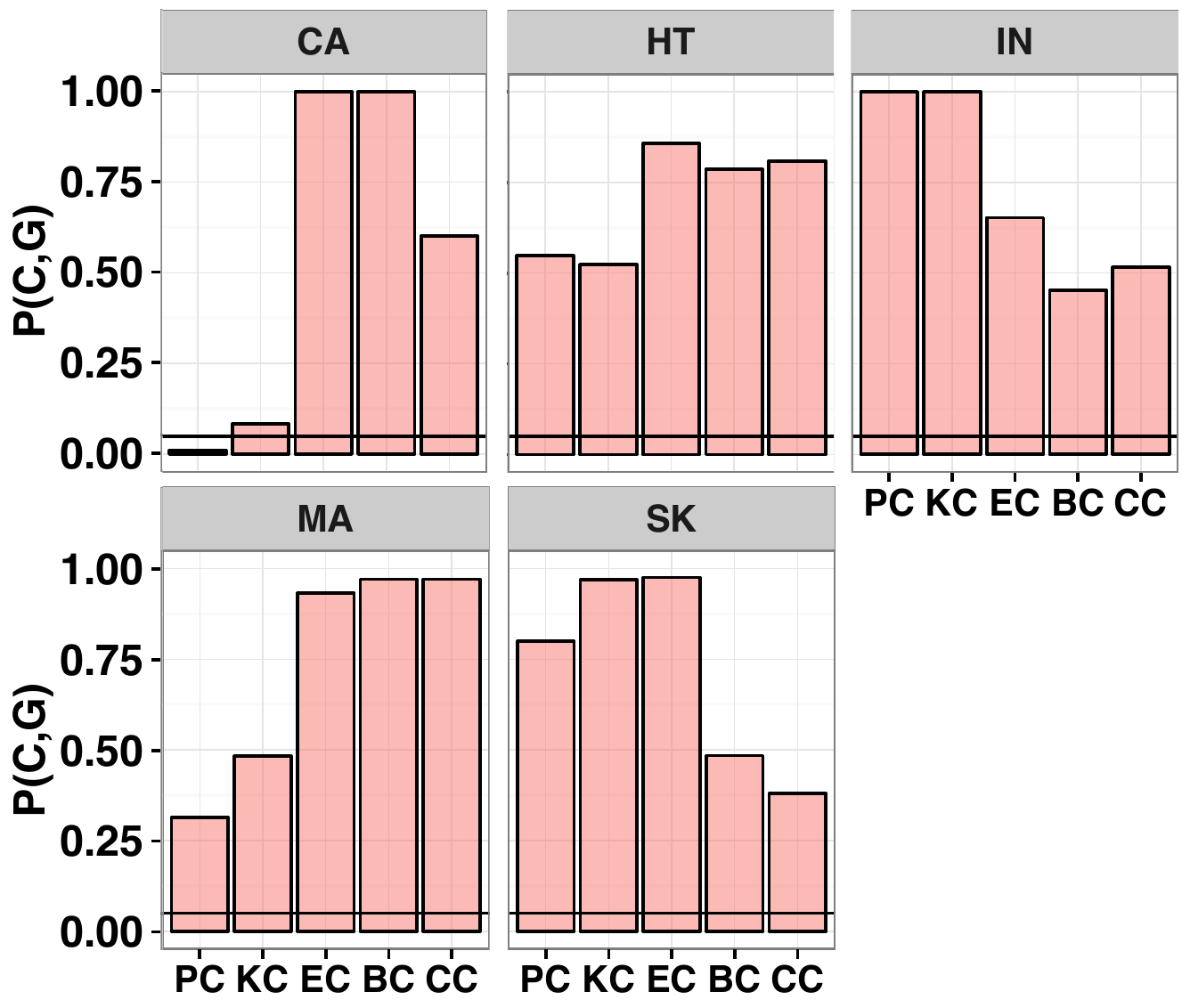}
\caption{Excerpt of results using the uniformly random null model used in the article}
\label{figure:origNull}
\end{figure}

To complete our study, Figure~\ref{figure:altNull} provides
the results of applying the methodology described in the article
using such an alternative null model. More precisely, we ran the same 
experiments where the null model is a random process that ensures
that the average degree of all sample graphs $H_i$ is the same as for $G_{t+1}$.
Figure~\ref{figure:origNull} recalls
the results we obtained for the uniformly random null model for
comparison.

For 4 out of 5 datasets, namely HT (Hypertext conference), IN (Infectious), 
MA (Manufacture mails) and SK (Souk cocktail), results obtained in both
cases are very similar. For all networks the dynamic signatures are strong, 
in the sense that the networks are outliers for many of the studied centralities
and the signatures of different networks vary, illustrating their unique 
evolution paths.
As expected, the ability of the presented method to 
distinguish the real network evolution compared to the networks generated 
according to the more refined null model decreases for most network traces and 
centralities. 

Yet, results are strikingly different from the more general null model in the 
main part of the paper for the case 
of CA, the Caida dataset. Caida differs from the other
datasets in the sense that it does not directly derive
from human activity (Caida captures Autonomous Systems relationships),
and the density in this dataset is much higher than in other
considered datasets, while the graph edit distance between different snapshots 
does not vary much.

\end{appendix}

{

}

\end{document}